\documentclass[12pt,a4paper]{article}

\usepackage{amsmath}
\usepackage{graphicx}
\usepackage{times}
\begin{document}
\begin{titlepage}
\title{Centrality in small systems at the LHC energies and beyond}
\author{ S.M. Troshin, N.E. Tyurin\\[1ex]
\small  \it NRC ``Kurchatov Institute''--IHEP\\
\small  \it Protvino, 142281, Russian Federation,\\
\small Sergey.Troshin@ihep.ru}
\normalsize
\date{}
\maketitle

\begin{abstract}
Experimental observations at the LHC of the collective effects in small systems such as $pp$--scattering
 suggest introduction of the centrality variable  to classify collision events  similar to the case of nucleus--nucleus collisions.  We discuss this issue
with reference to the existing measurements at the LHC and  the asymptotical energy region.
\end{abstract}
\end{titlepage}
\setcounter{page}{2}
\section{Introduction}
Start of the LHC operation for physics research  brought from  the very beginning an unexpected discovery of the collective effects  in the small systems \cite{cms} (for the comprehensive list of references to the experimental results of ALICE, ATLAS, CMS and LHCb Collaborations cf. \cite{mangano} and for a brief review --- \cite{trib}).  The most prominent effect is an observation of a "ridge" effect in two--particle  correlations in 
$pp$-collisions in the events with high multiplicity \cite{weili}. 
 The interest to the collective effects is supported due to their relation to mechanism of the quark-gluon plasma formation \cite{rafel}.  Earlier similar collective effects have been observed in the nuclei collisions  at RHIC.

The centrality  is  commonly accepted variable for description and classification of the collision events in nuclei interactions. This variable determines the degree of the collision peripherality. Thus, centrality is given by the impact parameter value attributed to the important geometrical characteristics of a particular  collision event. One should note that the impact parameter, is not a directly measurable quantity even in the nuclear collisions. 

This note is addressed to centrality determination for the small systems case such as $pp$--collisions.
 \section{Centrality in nucleus--nucleus collisions }
As it was noted in \cite{olli}, in experiments with nuclei the knowledge of centrality $c^N_b$ is inferred from either number of charged particles registered in the detector or the transverse energy measured in the calorimeter. Those quantities  are both denoted by $n$ and relevant to  experimentally measurable quantity   $c^N$ also called centrality.   Superscript $N$ means nuclear collisions, and in what follows, superscript $h$ will denote hadron collisions. The respective definitions of $c^N_b$ and $c^N$ have been borrowed from \cite{olli}. Namely,
\begin{equation}
\label{centb}
c_b^N\equiv \frac{1}{\sigma_{inel}}\int_0^bP^N_{inel}(b')2\pi b'db',
\end{equation}
where $P^N_{inel}(b)$ is the probability of an inelastic collision at the value of the impact parameter $b$, while the experimentally measurable quantity 
\begin{equation}
\label{cent}
c^N\equiv \int_n^\infty P^N(n')dn'.
\end{equation}
includes distribution over the multiplicity or the total transverse energy in the final state. 
It should be noted that the energy dependence of the above quantities is tacitly implied and not indicated explicitly. This energy dependence, however, can be a nontrivial one in the collisions of nuclei as well as of hadrons, since size of interaction region, probabilities of interactions, multiplicities and transverse energies are the energy--dependent quantities in both cases. Evidently, the effects related to the energy dependence of the above quantities should be taken into account under analysis of the data at the same centralities but different energies.

Under assumption that the probability $P^N(n)$ has a Gaussian distribution for a fixed value of the impact parameter $b$ \cite{bron},  the relation between  $c^N$ and $c_b^N$ has been obtained and discussed in \cite{olli}. This prescription allows one to get a precise knowledge on the impact parameter value from the experimental data. The assumption on the Gaussian distribution is a completely general one and does  not depend on the particular structure of the object under consideration, i.e. it can be applied for nuclei and hadrons as well.
 It is important to emphasize again  that the proposed reconstruction of the impact parameter is not based on a particular nuclear interaction model and/or concept of participating nucleons which numbers are usually obtained by a Glauber Monte Carlo simulation but not measured. 

\section{Centrality in small systems}
In view of the prominent collective effects observed in small systems, such as $pp$-collisions, it is natural to introduce centrality variable  for this case also to classify the collision events. 

The hadron scattering has similarities as well as differences with the scattering of nuclei.  Hadrons are also extended objects, but a significant contribution to $pp$--interactions is provided by the elastic scattering with the ratio of elastic to total cross-sections $\sigma_{el}(s)/\sigma_{tot}(s)$ currently rising with energy . The geometrical properties of hadron collisions revealed in elastic scattering   are essential   for  the hadron dynamics understanding, i.e. for the  development of QCD in its nonperturbative sector where the confinement plays a crucial role. The notion of the geometrical properties is relevant for the  interaction region of hadrons and not to the spacial properties of the individual hadrons. 

In what follows it will be argued that for the hadron interactions  definition of centrality by using a direct analogy with nuclei interaction  does not work. To give an appropriate definition, we  propose to use the full probability $P^h_{tot}(s,b)$ to take into account a contribution from the elastic channel. This contribution can be large and therefore should be taken into account.
Thus, for the centrality $c_b^h(s,b)$ the following definition is suggested
\begin{equation}
\label{centhb}
c_b^h(s,b)\equiv \frac{1}{\sigma_{tot}}\int_0^bP^h_{tot}(s,b')2\pi b'db',
\end{equation}
where
 $P^h_{tot}(s,b)\equiv \sum_{n=2}P^h_{n}(s,b) =\mbox{Im}f(s,b)$ and has a central impact parameter profile, i.e. its maximum is located at $b=0$.
The function $f(s,b)$ is the Fourier--Bessel transform of the scattering amplitude $F(s,t)$:
\begin{equation}\label{imp}
 F(s,t)=\frac{s}{\pi^2}\int_0^\infty bdbf(s,b)J_0(b\sqrt{-t}).
 \end{equation}
 The impact parameter representation diagonalises the unitarity relation for  the elastic scattering amplitude:
 \begin{equation}
 \mbox{Im} f(s,b)=|f(s,b)|^2+h_{inel}(s,b), \label{ub}
 \end{equation}
 where $h_{inel}(s,b)$ is the contribution of all intermediate inelastic channels. 
This representation provides a simple semiclassical picture of hadron scattering, i.e. head on or central collisions correspond to small impact parameter values.
For simplicity, we use a common assumption on the smallness of the
 real part of the elastic scattering amplitude $f(s,b)$   and perform replacement $f\to if$.

A comment on the global geometrical property of hadron interactions region should be made. 
It was shown in \cite{dips} that geometric (reflective) elastic scattering provides a leading contribution to 
$\langle b^2\rangle_{tot}(s)$ at $s\to\infty$ in case of unitarity saturation.
This conclusion emphasises necessity of the elastic channel accounting for determination of the hadron interaction region. 

To discuss the distinctive features of the absorptive and reflective scattering modes one should note first that
unitarity provides the following relation for the probability of the inelastic collisions $P^h_{inel}(s,b)$ in the case of proton--proton scattering
\begin{equation}\label{pinel}
P^h_{inel}(s,b) \equiv 4h_{inel}(s,b)= 4f(s,b)(1-f(s,b)).
\end{equation}
It  constraints variation of the  amplitude $f(s,b)$ by the values from the interval
$0 \leq f \leq 1$. The value of $f=1/2$ corresponds to the complete absorption of the initial state and means that the elastic scattering matrix element  is zero, $S=0$  (note that $S=1-2f$).  If the amplitude $f(s,b)$ at $b=0$ (beyond some threshold value of energy) becomes greater than $1/2$,  then the maximal probability of inelastic collisions has maximal value of $1$ at $b> 0$ (cf. Eq. (\ref{pinel})). 

 To demonstrate transition to the reflective scattering mode in a more clear way we use a particular unitarization scheme which represents the scattering amplitude $f(s,b)$ in the rational form and allows its variation in the whole interval allowed by unitarity \cite{umat}.
Respective form for the function $S(s,b)$ is written in this case as a known Cayley transform mapping nonnegative real numbers (it should be repeated here  that we neglect by the real part of the scattering amplitude $f(s,b)$)  to the interval $ [-1, 1]$:
\begin{equation}
S(s,b)=\frac{1-U(s,b)}{1+U(s,b)}. \label{umi}
\end{equation}
The real, nonnegative function $U(s,b)$ can be considered as the input or bare amplitude which is subject to the unitarization procedure. 
The models of a different kind can be used for construction of the function $U(s,b)$.
The most of the models provide increasing
dependence of the function $U(s,b)$ with energy (e.g. power-like one) and its exponential decrease with  the impact parameter due to analyticity in the Lehmann-Martin ellipse.
The value of the collision energy corresponding to the complete absorption of the initial state
at the central collisions  $S(s,b)|_{b=0}=0$
is $s_r$ and it is determined by the  equation
$U(s,b)|_{b=0}=1$\footnote{The estimates for the value of $s_r$ give $\sqrt{s_r}=2-3$ $TeV$ \cite{srvalue}. This is confirmed by the impact parameter analysis performed in \cite{alkin}.}.
In the energy region $s\leq s_r$ the scattering in the whole range of impact parameter variation
has a shadow nature. And when the energy becomes higher than 
 $s_r$, the scattering picture at small values of impact parameter 
($b\leq r(s)$, where  $S(s,b=r(s))=0$) starts to acquire a reflective contribution, when the
elastic $S$-matrix element varies in the region  $-1<S(s,b)\leq 0$ (at $s\geq s_r$). The term reflective is used by analogy with optics
since the phases of incoming state and outgoing  state differ by $\pi$.
The emerging physical picture of  very high energy scattering can be interpreted
as scattering off the reflecting
disk (approaching to complete reflection at the center) which is surrounded by a   black ring.
The reflection  mode leads to the limiting behavior $S(s,b)|_{b=0}\to -1$ at $s\to\infty$.

A wide class of the geometrical models (relevant for the centrality discussion) assume that $U(s,b)$ has a factorized form (cf. \cite{factor} and references therein):
\begin{equation}\label{usb}
U(s,b)=g(s)\omega(b),
\end{equation}
where $g(s)\sim s^\lambda$ at the large values of $s$, and the power dependence guarantees asymptotic growth of the total cross--section $\sigma_{tot}\sim \ln^2 s$. Such factorized form provides common origin for the increase of the total cross--sections and the slope of the diffraction cone in  elastic scattering. The particular simple form of the function $\omega(b)\sim \exp(-\mu b)$ has been chosen to meet the analytical properties of the scattering amplitude. 

Note, that only a weak energy dependence of the centrality variable provides a justified approach to the data analysis at the different energies.
Asymptotically, the centrality $c_b^h(s,b)$ defined according to Eq. (\ref{centhb}) decreases with energy slowly, like $1/ln^2(s)$ at fixed impact parameter values. It has the form 
$\pi b^2/\sigma_{tot}(s)$ since $f(s,b)$ saturates unitarity limit, $f(s,b)\to 1$ , at $s\to\infty$. Such slow decrease of centrality with energy would allow one to compare the data at different energy values at approximately the same value of centrality provided     the energy values are not too  different. Otherwise, there appear problems under comparison of the data obtained at the same values of centrality but at different energies.

The above indicated form of the function $\omega(b)$ can also be argued by the physical picture grounded on the
   represention of it as a convolution of the two energy--independent hadron pionic-type matter distributions in transverse plane \cite{chy}:
\begin{equation}
\omega (b)\sim D_1\otimes D_2\equiv \int d {\bf b}_1 D_1({\bf b}_1)D_2({\bf b}-{\bf b}_1).
\end{equation}
The function $\omega (b)$ reflects matter distributions (formfactors) of the constituents \cite{chiral}.

To justify additionally  the suggested form of Eq. (\ref{centhb}) for centrality, we examine Eq. (\ref{centb}) 
with \begin{equation}
\label{hinel}
h_{inel}(s,b)=\frac{U(s,b)}{[1+U(s,b)]^2}
\end{equation}
and the function $U(s,b)$ chosen according to Eq. (\ref{usb}) to calculate  explicitly the centrality given by Eq. (\ref{centb}). In this case: 
\begin{equation}
\label{cbh}
c_b^h(s,b) = \frac{1}{\ln(1+g(s))}\left[\ln\frac{1+g(s)}{1+g(s)\exp(-\mu b)}-\mu b f(s,b)\right],
\end{equation}
where 
\begin{equation}
\label{fsb}
f(s,b)=\frac{g(s)\exp(-\mu b)}{1+g(s)\exp(-\mu b)}.
\end{equation}
It allows one to conclude that at $s\to \infty$ and fixed value of $b$ one would expect strong energy decrease of the function $c_b^h(s,b)$, i.e.
\begin{equation}
\label{asymp}
c_b^h(s,b)\sim \frac{1}{s^\lambda\ln(s)}.
\end{equation}
This conclusion correlates with the statement made in \cite{hprod}, where it has been shown that centrality defined by a straightforward analogy with the nuclei collisions  cannot serve as a measure of the collision impact parameter but this quantity is to be associated with the dynamics of the multiparticle production when the value of impact parameter can variate in the narrow region arond $b=r(s)$. In this region absorption is maximal in case of the reflective scattering domination at $s\to\infty$.

The important problem is how to estimate experimentally   the impact parameter value in a particular $pp$-collision event. It is evident that the  event classification by multiplicity of the final particles  is not relevant for that purpose since a contribution of the elastic channel to the overall multiplicity is very small. Moreover, centrality defined in that way  has a strong energy dependence due to increasing peripherality of $P_{inel}^h(s,b)$ with energy. 

The most relevant observable seems to be a sum of the transverse energies of the final particles. We assume, following \cite{olli}, Gaussian distribution of the transverse energy for the fixed value of impact parameter $b$  implying the central limit theorem validity and extend the conclusion of \cite{olli} to $pp$--collisions.  Namely, fitting the experimental data and using Bayes' theorem for conditional probability one can reconstruct the distribution of the impact parameter for a given  value of centrality determined by the total transverse energy of all the final particles.
\section{Conclusion}
We have discussed a possible way of  centrality determination for the small systems like $pp$--collisions. We propose to use for that purpose transverse energy measurements as a tool for centrality assessments  in such systems instead of the multiplicity measurements.  The case of scattering with the saturation of unitarity is also included.  Nowadays, ATLAS and CMS experiments at the LHC are indeed using transverse energy measurements for centrality determination, but in the collisions of nuclei only. Further implication of this proposal would include discussion of its  experimental feasibility.
\section*{Acknowledgements}
We are grateful to E.S. Martynov for the interesting correspondence on the problem of centrality  in small systems.

\end{document}